\documentstyle[11pt,paspconf]{article}
\def\edcomment#1{\iffalse\marginpar{\raggedright\sl#1\/}\else\relax\fi}
\reversemarginpar

\begin{document}
\title{Bulk Motions in Large-Scale Void Models}
\author{Kenji Tomita}
\affil{Yukawa Institute for Theoretical Physics, Kyoto University,
        Kyoto 606-8502, Japan}

\begin{abstract}
To explain the puzzling situation in the observed bulk flows on scales 
$\sim 150 h^{-1}$ Mpc \ ($H_0 = 100 h^{-1}$ km sec$^{-1}$
Mpc$^{-1}$), we consider 
the observational behavior of spherically symmetric inhomogeneous
cosmological models, which consist of inner and
outer homogeneous regions connected by a shell or an intermediate
self-similar region. It is assumed that the present matter density
parameter in the inner region is smaller than that in the outer
region, and the present Hubble parameter in the inner region is 
larger than that in the outer region. Then galaxies in the inner 
void-like region can be seen to have a bulk motion 
relative to matter in the outer region, when we observe them at a
point O deviated from the center C of the inner region. Their velocity
$v_p$ in the CD direction is equal to the difference of two Hubble 
parameters multiplied by the distance between C and O.
It is found also that the velocity $v_d$ corresponding to CMB dipole 
anisotropy observed at O is by a factor $\approx 10$ small compared with
$v_p$. This behavior of
$v_d$ and $v_p$ is consistent with the observed cosmic flow
of cluster galaxies, when the radius of the inner region and the
distance CD are about 200 $h^{-1}$ Mpc and 40 $h^{-1}$ Mpc,
respectively, and when the gaps of density and Hubble parameters 
are $\approx 0.5$ and $18 \%$, respectively. Moreover, the [$m, z$]
relation in these models is discussed in connection with SNIa data.  

\end{abstract}

\section{Introduction}

The dipole moment in the cosmic background radiation (CMB) is thought
to come mainly from the Doppler shift due to the motion of the Local
Group (LG), relative to the cosmic homogeneous expansion. As the main
gravitational source which brings the velocity vector of LG, the
existence of the Great Attractor (GA) was found by Lynden-Bell et
al.(1988) and Dressler et al. (1987). It has the
position at the redshift of 4300 km sec$^{-1}$. On the other hand, the 
motion of LG relative to the inertial frame consisting of many clusters on
larger-scales was studied observationally by several groups: A
bulk flow of $\sim 700$ km sec$^{-1}$ was found by Lauer and Postman
(1994, 1995) and Colless (1995) as the motion of the Abell cluster 
inertial frame relative to LG in the region with redshift $< 15000$ 
km sec$^{-1}$, but in the other approach the different result was 
derived by Giovanelli et al. (1998), Dale et al. (1999) and Riess et
al. (1997) in the regions with similar redshifts.
Lauer and Postman's work is based on the assumption that the
brightest cluster galaxies as standard candles and the Hoessel
relation can be used, but at present these assumptions have been
regarded as questionable or unreliable.

Independently of these works, the motion of cluster frames relative to 
CMB was measured by Husdon et al. (1999) and Willick (1999) due to the global 
Hubble formula
using the Tully-Fisher distances of clusters and their redshifts with
respect to CMB, and the flow velocity vector was derived in the region 
with about 150$h^{-1}$ Mpc  \ ($H_0 = 100 h^{-1}$ km sec$^{-1}$ 
Mpc$^{-1}$).  The remarkable and puzzling properties of these flows 
are that the flow velocity
reaches a large value $\sim 700$ km/sec on a large scale, while the
dipole velocity (not due to GA) corresponding to the CMB dipole 
anisotropy seems to be much small, compared with the above flow velocity. 

In the present note we first consider inhomogeneous models on
sub-horizon scale, corresponding to matter flows on scales $\sim 150 h^{-1}$
Mpc. They are assumed to be spherically symmetric
inhomogeneous models which
consist of inner and outer homogeneous regions connected by a shell
being a singular layer, and the behavior of large-scale motions
caused in the inner region is considered. Next we consider light rays 
which are emitted at the last scattering surface and reach an observer 
situated at a point O (inthe inner region) deviated from the center C, 
and the CMB dipole anisotropy for the observer is shown.
On the basis of these results we show the consistency with various 
observations of cosmic flows.  
Moreover the [$m, z$] relation is discussed in connection with SNIa
data, and finally concluding remarks are presented.

\section{Cosmological models and the bulk motions}

In previous papers (Tomita 1995, 1996) we treated spherically 
symmetric inhomogeneous models which consist of inner and outer 
homogeneous regions connected with an intermediate self-similar 
region and have the boundary on a super-horizon scale.     
Here we consider a similar spherically symmetric inhomogeneous model which
consists of inner and outer homogeneous regions, but is connected by a shell
being a singular layer on a sub-horizon scale $\sim 150 h^{-1}$ Mpc.
This shell may be associated with large-scale structures or excess 
powers observed by Broadhurst et al.(1990), Landy et al.(1996), and 
Einasto et al.(1997).
The physical state in each region is specified by the Hubble constant
density and the density parameter. It is assumed that   
 the present Hubble parameter in the inner region ($H_0^{\rm in}$)
is larger than that in the outer region ($H_0^{\rm out}$), and the 
present inner density parameter ($\Omega_0^{\rm in}$)
is smaller than the present outer density parameter ($\Omega_0^{\rm out}$).
The evolution of physical states in each region and the boundary has been 
studied in the form of void models (e.g., Sakai et al. 1993).

The average motion of CMB is comoving with matter in the outer region,
while it is not comoving with matter in the inner region or matter in
the inner region moves relative to CMB, because their Hubble constants 
are different. The bulk motion appears as the result of this relative
motion to CMB. The relative velocity ($\Delta v$) is \ $(H_0^{\rm in} -
H_0^{\rm out}) r$ \ in the radial direction, where $r$ is the radial
distance from the center C to an arbitrary point (a cluster's
position) in the inner region. When an observer O sees this velocity 
vector $\Delta v$, it can be divided into two parts: the component in the
observer's line of sight ($\Delta v_{\rm ls}$) and the bulk-velocity
 component in the direction of C $\rightarrow$ O ($v_p$). The latter
component $v_p$ is constant, irrespective of the cluster's
position. In the case when the present radius of the boundary and the
observers's position are $\sim 200 h^{-1}$ and $40 h^{-1}$ Mpc,
respectively, we have $v_p \sim 700$ km sec$^{-1}$.

\section{Dipole anisotropy and the consistency with
various observations of cosmic flows}

If the observer were in the center C, he never sees any CMB
anisotropy, as long as the two regions are homogeneous. For the
non-central observer O we have nonzero dipole anisotropy $D$ which is
derived by calculating curved paths from the last scattering surface
to O and the directional variation of the temperature $T_r$. The
velocity $v_d$ corresponding to $D$ is defined by
$v_d \equiv c [(3/4\pi)^{1/2} D]$ and derived.
As the result it was found that $v_d$ is small compared with $v_p$, if 
O is near to C. In our above example, $r({\rm OC}) / r({\rm boundary}) 
\sim 1/5$, we obtain $v_d \sim 0.1 v_p$.

As described in \S 2, the bulk velocities at arbitrary two points are
equal and so their difference is zero. Accordingly the relative velocity 
of the Local Group (LG) to the frame of clusters ($v_{LG}$) is only
the peculiar velocity ($v_{GA}$) caused by the small-scale
nonspherical gravitational field of the Great Attracter. The above
result gives the dipole velocity of LG, $v_d ({\rm LG}) = v_{GA} +
v_d$, so that $v_{GA}$ and $v_d ({\rm LG})$ are comparable and the
diffrence is $v_d \ (\sim 0.1 v_p)$.  
 This situation in the present models is consistent with the
observations (Giovanelli et al. (1998), Dale et al. (1999) and Riess
et al. (1997)) for relative velocities of LG to the cluster frame     
, and the observations (Husdon et al. (1999) and Willick (1999)) for
the bulk flows of clusters, since the observed values of $v_{LG}$,
$v_d ({\rm LG})$ and $v_p$ are  $565$ km sec$^{-1}$, $627$ km sec$^{-1}$
and $\sim 700$ km sec$^{-1}$, respectively, in the similar directions. 
The observed difference of first two velocities is about $0.1 \times v_p$.

The detail derivation of the contents in \S 2 and \S 3 is shown in
Tomita (1999a).  
  
\section{[$m, z$] relation and SNIa data}

Here the behavior of distances in the present models is studied.
First we treat the distances from a virtual observer who is in 
the center C of the inner void-like region in models with a single
shell, and derive the 
[magnitude $m$ - redshift $z$] relation. This relation is compared
with the counterpart in the homogeneous models. Then the relation 
in the present models is found to deviate from that in the homogeneous
models with $\Lambda = 0$ at the stage of $z < 1.5$. It is partially 
similar to that in the nonzero-$\Lambda$ homogeneous models,
but the remarkable difference appears at the high-redshift stage $z > 1.0$. 
Moreover, we consider a realistic observer who is in the
position O deviated from the center, and calculate the distances from
him. The distances depend on the direction of incident light and the
area angular diameter distance is different from the linear angular diameter
distances. It is shown as the result that the [$m, z$] relation is
anisotropic, but the relation averaged with respect to the angle is
very near to the relation by the virtual observer. When we compare
these theoretical relations with SNIa data (Riess et al.(1998),
Garnavich et al.(1998), and Schmidt et al.(1998)), we can determine 
which of
the present models and nonzero-$\Lambda$  homogeneous models are
better, and the fittest model parameters. At present, however, there 
are few data at $z \sim 1.0$, so that the model selection may not be
performed.  The detail description of the content in this section is
given in Tomita (1999b). 

\section{Concluding remarks}
The density perturbations in the inner region and their influence of 
 on CMB anisotropy are another important factor to the selection of
 model parameters, which should be studied next.


\begin{references}
\reference Lynden-Bell, D., Faber, S.
M., Burstein, D., Davies, R. L., Dressler, A., Terlevich, R. J. and
Wegner, G. 1988, \apj, 326, 19
\reference Dressler, A., Faber, S.M., Burstein,
D., Davies, R. L., Lynden-Bell, D., Terlevich, R. J., and
Wegner, G. 1988, \apj, 313, L37
\reference Lauer, T. R. and Postman, M. 
1994, \apj, 425, 418;  Postman, M. and Lauer, T. R. 1995, \apj, 440, 28
\reference Colless, M. 1995, \aj, 109, 1937
\reference Giovanelli, R., Haynes,
M. P., Freudling, W., da Costa, L. N., Salzer, J. J., and Wegner,
G. 1998, \apj, 505, L91  
\reference Dale, D. A., Giovanelli, R., and
Haynes, M.P. 1999, \apj, 510, L11
\reference Riess, A. G., Davis, M., Baker,
J., and Kirshner, R. P. 1997, \apj, 488, L1 
\reference Hudson, M. J., Smith, R. J., 
Lucey, J. R., Schlegel D. J. and Davies, R. L. \apj, 512, L79
\reference Willick, J. A. 1999, \apj, 516, 47
\reference Tomita, K. 1996, \apj, 461, 507
\reference Tomita, K. 1995, \apj, 451, 541;
 Tomita, K. 1996, \apj, 464, 1054   for Erratum
\reference Broadhurst, T. J., Ellis,
R. S., Koo, D.C., Szalay, A. S. 1990, Nature, 343, 726
\reference Landy, S. D., Shectman, S. A.,
Lin, H., Kirshner, R. P., Oemler, A., and Tucker, D. 1996, \apj, 456, L1 
\reference Einasto, J. et al. 1997, Nature, 385, 139
\reference Sakai, N.,  Maeda, K. 
and Sato, H. 1993, Prog. Theor. Phys., 89, 1193 
\reference Tomita, K. 1999a, \apj, in press (astro-ph/9905278)
\reference  Riess, A. G. et al. 1998, \aj, 116, 1009
\reference  Garnavich, P. M. et al., 1998, \apj, 493, L53
\reference  Schmidt, B. P. et al. 1998, \apj, 507, 46
\reference Tomita, K. 1999b, \apj, in press (astro-ph/9906027)
\end{references}
\end{document}